\documentstyle{mn}

\input epsf

\begin{document}

\def\Lya{Ly$\alpha\ $}
\def\DI{\hbox{D~$\scriptstyle\rm I\ $}}
\def\HI{\hbox{H~$\scriptstyle\rm I\ $}}
\def\bHI{\hbox{\bf H~$\scriptstyle\bf I\ $}}
\def\HII{\hbox{H~$\scriptstyle\rm II\ $}}
\def\HeI{\hbox{He~$\scriptstyle\rm I\ $}}
\def\HeII{\hbox{He~$\scriptstyle\rm II\ $}}
\def\HeIII{\hbox{He~$\scriptstyle\rm III\ $}}
\def\OIII{\hbox{O~$\scriptstyle\rm III\ $}}
\def\kms{\,{\rm {km\, s^{-1}}}}
\def\msun{{M_\odot}}
\def\nHI{{\rm HI}}
\def\nH{{\rm H}}
\def\nHII{{\rm HII}}
\def\mpc{\,{\rm {Mpc}}}
\def\kmsmpc{\,{\rm km\,s^{-1}\,Mpc^{-1}}}
\def\hmpc{\,h^{-1}{\rm \,Mpc}\,}
\def\mpch{\,h{\rm \,Mpc}^{-1}\,}
\def\hkpc{\,h^{-1}{\rm \,kpc}\,}
\def\ev{\,{\rm eV\ }}
\def\kel{\,{\rm K\ }}
\def\intunits{\,{\rm ergs\,s^{-1}\,cm^{-2}\,Hz^{-1}\,sr^{-1}}}
\def\ltsima{$\; \buildrel < \over \sim \;$}
\def\lsim{\lower.5ex\hbox{\ltsima}}
\def\gtsima{$\; \buildrel > \over \sim \;$}
\def\gsim{\lower.5ex\hbox{\gtsima}}
\def\etal{{ et~al.~}}
\def\aj{AJ}
\def\ana{A\&A}
\def\apj{ApJ}
\def\apjs{ApJS}
\def\mn{MNRAS}
\def\spose#1{\hbox to 0pt{#1\hss}}
\def\lta{\mathrel{\spose{\lower 3pt\hbox{$\mathchar"218$}}
     \raise 2.0pt\hbox{$\mathchar"13C$}}}
\def\gta{\mathrel{\spose{\lower 3pt\hbox{$\mathchar"218$}}
     \raise 2.0pt\hbox{$\mathchar"13E$}}}

\journal{Preprint-00}

\title{Energy transfer by the scattering of resonant photons}

\author[A. Meiksin]{Avery Meiksin \\
SUPA\thanks{Scottish Universities Physics Alliance},
Institute for Astronomy, University of Edinburgh,
Blackford Hill, Edinburgh\ EH9\ 3HJ, UK}


\maketitle

\begin{abstract}
A formal derivation is presented of the energy transfer rate between
radiation and matter due to the scattering of an isotropic
distribution of resonant photons. The derivation is developed in the
context of the two-level atom in the absence of collisions
and radiative transitions to and from the continuum, but
includes the full angle-averaged redistribution function for photon
scattering. The result is compared with previous derivations, all of
which have been based on the Fokker-Planck approximation to the
radiative transfer equation. A new Fokker-Planck approximation,
including an extension to higher (post-diffusive) orders, is derived
to solve the radiative transfer equation, and time-dependent numerical
solutions are found. The relaxation of the colour temperature to the
matter temperature is computed as the radiation field approaches
statistical equilibrium through scattering.  The results are discussed
in the context of the Wouthuysen-Field mechanism for coupling the 21cm
spin temperature of neutral hydrogen to the kinetic temperature of the
gas through \Lya scattering. The evolution of the heating rate is also
computed, and shown to diminish as the gas approaches statistical
equilibrium.
\end{abstract}


\section{Introduction}

The prospects for detecting the Intergalactic Medium (IGM) prior to
reionization by extremely large radio telescopes like the {\it Low
Frequency Array} (LOFAR) and the {\it Square Kilometre Array} (SKA)
has led to a renewed interest the Wouthuysen-Field mechanism for
coupling the spin temperature of neutral hydrogen to the kinetic
temperature of the gas through the scattering of \Lya photons
(Wouthuysen 1952; Field 1958). Following Field's (Field 1959a, b),
analysis of the scattering of \Lya photons in a homogeneous medium and
their role in setting the spin temperature of neutral hydrogen (Field
1958), the physics of 21cm excitation and absorption was re-examined
in the light of current cosmological models by Madau, Meiksin, \& Rees
(1997, hereafter MMR) and Tozzi \etal (2000). A large number of
various scenarios and effects have since been explored in the
literature (see Bowman, Morales \& Hewett 2005 and references
therein).

In addition to de-coupling the hydrogen spin temperature from the
temperature of the Cosmic Microwave Background (CMB), MMR showed that
\Lya photons were in principle capable of heating cold hydrogen gas
through recoils. The derivation was based on the momentum transferred
to the hydrogen atoms by the \Lya photons after recoil (Field 1959b,
Basko 1981). MMR supported their argument by identifying the recoil
heating term in the radiative transfer equation for the scattering of
the photons using a Fokker-Planck approximation with atomic recoil of
Rybicki \& Dell'Antonio (1994), who extended previous Fokker-Planck
treatments (Unno 1955; Harrington 1973; Basko 1981; Krolik 1990) to
include the full Voigt line profile. While recognising that the energy
exchange would cease once the radiation and matter reached thermal
equilibrium, MMR did not discuss the role of photon diffusion through
frequency on the energy exchange rate, as its effect on energy
exchange is negligible for sufficiently cold gas. As the radiation
field approaches statistical equilibrium through scatterings, however,
the photon diffusion term reduces the net heating rate, which
ultimately vanishes once the radiation field achieves statistical
equilibrium at frequencies near resonance, with a colour temperature
that matches the matter temperature. This will generally occur well
before the radiation field reaches thermal equilibrium with the
matter. The degree to which the approach to statistical equilibrium
reduces the heating rate depends on the frequency range over which
equilibrium is achieved. While it may be achieved in the core of the
absorption profile on a timescale of the order of the mean free time
for scattering at line-centre, it will take longer outside the core
and significant heating may still persist. A central purpose of this
paper is to derive the energy exchange rate from the full radiative
transfer equation, justify the use of the Fokker-Planck equation for
describing the exchange of energy between the radiation and the
matter, and compute the evolution of the heating rate in this
approximation for some simple applications. This is done in the
approximation of the two-level atom, neglecting collisions and
scatterings to and from the continuum. A second purpose it to compute
the timescale over which the colour temperature relaxes to the kinetic
temperature of the gas, as this is crucial for determining the
signature of 21cm emission or absorption through the Wouthuysen-Field
mechanism.

Under the assumption that the Fokker-Planck equation of Rybicki \&
Dell'Antonio (1994) adequately described the exchange of energy with
the matter, Chen \& Miralda-Escud\'e (2004) applied Rybicki \&
Dell'Antonio's cosmological version of the Fokker-Planck equation to
the heating of the IGM, assuming that the comoving radiation density
had reached a steady-state. In this case, the energy exchange between
the radiation field and the matter virtually ceases, as in the static
case, except for a residual amount arising as photons are redshifted
through the resonance frequency. No estimate, however, is made of the
time required to reach the equilibrium heating rate.

Following Deguchi \& Watson's (1985) formulation of the resonant
photon scattering problem including stimulated emission, Rybicki
(2006) presents an alternative derivation of the energy exchange rate
using a Fokker-Planck approximation, and derives an approximate form
for the radiative transfer equation that is structurally similar to
the Kompaneets equation.

In order to solve the radiative transfer equation, a different
Fokker-Planck approach is adopted here. The Fokker-Planck
approximation derived is more complete than any previously formulated
in this context. In particular, the formulation presented here
explicitly conserves photon number and energy to second order in
derivatives of the radiation field, as well as when extended to higher
orders, permitting more accurate solutions to be obtained in a
self-consistent manner than was previously possible. This remedies a
shortcoming of the Fokker-Planck treatment of Rybicki \& Dell'Antonio
(1994), for which particle conservation was imposed by hand. Since
then Rybicki (2006)
\footnote{The present paper was substantially complete, but not
submitted, prior to the posting of Rybicki (2006). Differences between
the present work and Rybicki's are commented on in the text.} has
presented an improved Fokker-Planck treatment. There are, however,
some differences between the Fokker-Planck treatment provided here and
Rybicki's. These are discussed below.

In the next section, the basic theory of resonance scattering is
reviewed, including a derivation of the rate of energy exchange
between the radiation field and the scattering medium. The
Fokker-Planck approximation is discussed in \S3, and applied to some
simple problems in \S4. A discussion of the results and conclusions
are presented in \S5.

\section{Basic theory}
\subsection{Radiative transfer and heating equations}
Statements of the radiative transfer equation in the literature are
generally approximations, and not all are adequate for formulating the
full energy transfer problem between the photons and the scattering
medium. A standard form of the radiative transfer equation in terms of
the photon energy density (Mihalas 1978, \S2) in fact leads to the
complete absence of energy transfer due to the implicit approximations
made. In this section, a derivation of the radiative transfer equation
satisfactory for treating the effects of energy exchange by the
scatter of resonant photons is presented. A formal expression for the
energy exchange rate for an isotropic radiation field is derived in
this section which uses the full scattering integral, unlike all
previous discussions which were based on the Fokker-Planck
approximation.

In order to clarify the assumptions made in the radiative transfer
equation for resonant photons, a derivation from basic principles is
provided. The problem is formulated in a non-relativistic
context. Polarisation is suppressed as well, but the results are
readily generalised.

In a static medium, the rate of change of the energy in the radiation
field at frequency $\nu$ at position ${\bf r}$ at time $t$ moving (in
the laboratory frame) into direction ${\bf \hat n}$ per unit area per
unit solid angle per unit time per unit frequency is (Mihalas 1978,
Section 2.2)
\begin{eqnarray}
\frac{D I_\nu({\bf r},t,{\bf \hat n})}{D t}
&=&\frac{\partial I_\nu({\bf r},t,{\bf \hat n})}{\partial t}+
c{\bf \nabla} I_\nu({\bf r},t,{\bf \hat n})\nonumber\\
&=&c\eta_\nu({\bf r},{\bf \hat n},t)-
c\chi_\nu({\bf r},{\bf \hat n},t)I_\nu({\bf r},t,{\bf \hat n}),
\label{eq:RTgen}
\end{eqnarray}
where $I_\nu({\bf r},t,{\bf \hat n})$ is the specific intensity,
$\eta_\nu$ is the specific emissivity (energy emitted per unit volume
per unit solid angle per unit time per unit frequency),
$\chi_\nu$ is the specific extinction (fraction of incident energy
removed per unit length), and $c$ is the speed of light. (Finite
light travel-time effects are neglected.)

For simplicity, it will be assumed that the emissivity and extinction
both arise entirely due to scattering by a two-level atom. (Much of
the following discussion is based on Mihalas 1978, \S13-4. See also
Milkey \& Mihalas 1973.) Scatterings to and from the continuum levels
are neglected, as are excitations and ionizations due to
collisions. Local isotropy of the radiation field is also assumed,
although global variations with position are allowed. Under these
assumptions, the rate of change of atoms in the upper state due to
interactions involving the production of photons of energy $h\nu$ is
given by
\begin{eqnarray}
\frac{d n_u(\nu)}{dt}&=&-n_u(\nu)\left(A_{ul}+B_{ul}J_\nu\right)\nonumber\\
&+&n_l B_{lu}\int_0^\infty d\nu'\,R(\nu',\nu)J_{\nu'},
\label{eq:dndt}
\end{eqnarray}
where $J_\nu=\oint(d\omega/4\pi)\, I_\nu$ is the usual angle-averaged
intensity. ($J_\nu=I_\nu$ for an isotropic radiation field.) The terms
on the right hand side correspond, respectively, to spontaneous decays
given by the Einstein coefficient $A_{ul}$, stimulated emissions
described by the downward coefficient $B_{ul}$ and scatterings from
the lower to upper level described by the upward coefficient
$B_{lu}$. The scattering function $R(\nu',\nu)$ is the angle-averaged
probability that a photon of frequency $\nu'$ is scattered to
frequency $\nu$ by an atom in the lower state. It is normalised by
$\varphi(\nu)=\int_0^\infty d\nu'\,R(\nu,\nu')$, where $\varphi(\nu)$
is the absorption profile of the atoms in the lower state. The
function $n_u(\nu)d\nu$ describes the number density of atoms in the
upper state capable of emitting photons in the frequency range
$(\nu,\nu+d\nu)$ in the observer's frame. In general $n_u(\nu)/n_u$,
where $n_u$ is the total number density of photons in the upper level,
is not given by the natural emission profile $\psi(\nu)=\int_0^\infty
d\nu'\, R(\nu',\nu)$ and must be solved for.

The emissivity is then given by
\begin{equation}
\eta_\nu=h\nu n_u(\nu)\frac{A_{ul}}{4\pi},
\label{eq:etanu}
\end{equation}
and the absorption coefficient by
\begin{equation}
\chi_\nu=h\nu \left[n_l \frac{B_{lu}}{4\pi}\phi_\nu-n_u(\nu)\frac{B_{ul}}
{4\pi}\right].
\label{eq:chinu}
\end{equation}
Expressed in terms of the radiation energy density $u_\nu=4\pi J_\nu/c$,
the radiative transfer equation becomes
\begin{eqnarray}
\frac{D u_\nu({\bf r},t)}{D t}
&=&\frac{\partial u_\nu({\bf r},t)}{\partial t}+
c{\bf \nabla} u_\nu({\bf r},t)\nonumber\\
&=&h\nu n_u(\nu)A_{ul}
-h\nu n_l\frac{B_{lu}}{4\pi}c\varphi(\nu)u_\nu\nonumber\\
&&\times\left[1-\frac{B_{ul}}{B_{lu}}
\frac{n_u(\nu)}{n_l\varphi(\nu)}\right].
\label{eq:dunudt}
\end{eqnarray}

In most applications, the level populations will have reached
equilibrium. Eq.~(\ref{eq:dndt}) then gives
\begin{equation}
n_u(\nu)=n_l\frac{B_{lu}\int d\nu'\,R(\nu',\nu)u_{\nu'}}
{4\pi A_{ul}/ c + B_{ul}u_\nu}.
\label{eq:nudb}
\end{equation}
Noting that the photon occupation number is ${\mathcal N}(\nu)=
c^3u_\nu/(8\pi h\nu^3)$, the rate equation for $u_\nu$ may be
expressed as
\begin{eqnarray}
\frac{D u_\nu({\bf r},t)}{D t}
&=&\frac{\partial u_\nu({\bf r},t)}{\partial t}+
c{\bf \nabla} u_\nu({\bf r},t)\\
&=&h\nu n_l({\bf r},t) c\frac{B_{lu}}{4\pi}
\frac{\int d\nu' R(\nu',\nu)u_{\nu'}({\bf r},t)}
{1+{\mathcal N}(\nu,{\bf r},t)}\nonumber\\
&&-h\nu n_l({\bf r},t)c\frac{B_{lu}}{4\pi}\varphi(\nu)u_\nu\nonumber\\
&\times&\left\{1-\frac{\int d\nu'\,R(\nu',\nu)
\left(\frac{\nu'}{\nu}\right)^3{\mathcal N}(\nu',{\bf r},t)}{\varphi(\nu)
\left[1+{\mathcal N}(\nu,{\bf r},t)\right]}
\right\},\nonumber
\label{eq:dunudtse}
\end{eqnarray}
after using the relations between the Einstein coefficients.
When the radiation reaches thermodynamic equilibrium, it becomes
blackbody radiation at the same temperature as the matter, so that
$Du_\nu/Dt=0$. This permits a relation to be derived between $R(\nu,\nu')$
and $R(\nu',\nu)$ in detailed balance. Setting
${\mathcal N}(\nu)=1/[\exp(h\nu/kT)-1]$ gives
\begin{equation}
\frac{R(\nu',\nu)}{R(\nu,\nu')}=\left(\frac{\nu}{\nu'}\right)^3
\frac{e^{h\nu'/kT}-1}{e^{h\nu/kT}-1}\rightarrow \left(\frac{\nu}{\nu'}\right)^3
e^{-h(\nu-\nu')/kT},
\label{eq:Rdetbal}
\end{equation}
where the limit applies for $h\nu>>kT$ and $h\nu'>>kT$ corresponding
to the classical limit of a dilute radiation field. This limiting form
corresponds to the result of Deguchi \& Watson (1985) (noting the
change in their normalization of $R(\nu,\nu')$). They claim their
result is general, although it seems to apply only for low values of the
photon occupancy.

Eq.~(7) simplifies. Defining the coefficient $\chi_0({\bf r},t)=
n_l({\bf r},t)(B_{lu}/4\pi)h\nu_0$, where $\nu_0$ is the resonant
frequency, after simplification the equation of radiative transfer
becomes
\begin{eqnarray}
\frac{D u_\nu({\bf r},t)}{D t}
&=&\frac{\partial u_\nu({\bf r},t)}{\partial t}+
c{\bf \nabla} u_\nu({\bf r},t)\nonumber\\
&=&\frac{\chi_0({\bf r},t)c}
{h\nu_0}h\nu\biggl[\int_0^\infty d\nu' R(\nu',\nu)u_{\nu'}\nonumber\\
\phantom{\biggl[}&&-\varphi(\nu) u_\nu({\bf r},t)\biggr]
+ h\nu S(\nu).
\label{eq:unuev}
\end{eqnarray}
A source term $S(\nu)$ of photons (per
unit volume) has also been added (which may be taken to be an
arbitrary function over space and time). It is noted that in the
presence of additional interactions that are capable of redistributing
the level populations, such as collisions or ionizations and
recombinations to and from the continuum, the relations become
considerably more involved. In general it is necessary to solve for
the detailed level populations including the emission profile in order
to establish the fraction of atoms in the lower and upper states. (See
Mihalas 1978 \S13-4 for details.)

An alternative form of the radiative transfer equation in terms of the
number density of photons $n_\nu=u_\nu/h\nu$ is given by dividing
Eq.~(\ref{eq:unuev}) by $h\nu$
\begin{eqnarray}
\frac{D n_\nu({\bf r},t)}{D t}
&=&\frac{\partial n_\nu({\bf r},t)}{\partial t}+
c{\bf \nabla} n_\nu({\bf r},t)\nonumber\\
&=&\frac{\chi_0({\bf r},t)c}
{h\nu_0}\biggl[\int_0^\infty d\nu' R(\nu',\nu)u_{\nu'}\nonumber\\
\phantom{\biggl[}&&-\varphi(\nu) u_\nu({\bf r},t)\biggr] + S(\nu).
\label{eq:nnuev}
\end{eqnarray}

The rate at which energy is transferred to the matter per unit volume
is given by integrating Eq.~(\ref{eq:unuev}) over frequency
\begin{eqnarray}
G({\bf r},t)&=&\int_0^\infty d\nu h\nu S(\nu)-\frac{D}{Dt}
\int_0^\infty d\nu u_\nu({\bf r},t)\\
&=&-\frac{\chi_0({\bf r},t)c}{h\nu_0}
\biggl[\int_0^\infty d\nu \int_0^\infty d\nu' h\nu
R(\nu',\nu)u_{\nu'}({\bf r},t)\nonumber\\
&&\phantom{\biggl[}\mbox{}-\int_0^\infty d\nu h\nu \varphi(\nu)
u_\nu({\bf r},t)\biggr]\nonumber\\
&=&\frac{\chi_0({\bf r},t)c}{h\nu_0}
\int_0^\infty d\nu \int_0^\infty d\nu' (h\nu'-h\nu)
R(\nu',\nu)u_{\nu'}({\bf r},t)\nonumber,
\label{eq:Gheat}
\end{eqnarray}
after noting that $\int d\nu R(\nu',\nu)=\varphi(\nu')$. The final form
shows that the energy transfer rate to the scattering medium is given
by the rate at which the photon energies in the observer's frame shift
during scattering.

Eqs.~(\ref{eq:unuev}) and (11) have some noteworthy
symmetry properties that may be used to gain insight into the origin
of the energy exchange in special situations. For resonant
scattering, the redistribution function $R(\nu',\nu)$ for most
applications is well-approximated as a function of
$(\nu-\nu')$. Similarly $\varphi(\nu)$ is an (even) function of
$(\nu-\nu_0)$. For a source function $S_\nu$ that is a function of
$(\nu-\nu_0)$, in the absence of an initial radiation field, the form
of Eq.~(\ref{eq:unuev}) shows that the radiation energy density
$u_\nu$ may also be expressed as a function $u(\nu-\nu_0)$. It is then
possible to express $R(\nu,\nu')$ and $u(\nu-\nu_0)$ as sums of their
even and odd parts in $(\nu-\nu_0)$, $R(\nu,\nu')=R^{(e)}(\nu,\nu') +
R^{(o)}(\nu,\nu')$ and
$u(\nu-\nu_0)=u^{(e)}(\nu-\nu_0)+u^{(o)}(\nu-\nu_0)$, where
$R^{(e)}(\nu,\nu')=[R(\nu,\nu') + R(\nu',\nu)]/2$,
$R^{(o)}(\nu,\nu')=[R(\nu,\nu') - R(\nu',\nu)]/2$,
$u^{(e)}(\nu-\nu_0)=[u(\nu-\nu_0)+u(\nu_0-\nu)]/2$ and
$u^{(o)}(\nu-\nu_0)=[u(\nu-\nu_0)-u(\nu_0-\nu)]/2$. The form of
Eq.~(11) then shows that energy exchange between the
radiation field and the matter is driven only by the even-odd
combinations $R^{(e)}u^{(o)}$ and $R^{(o)}u^{(e)}$. In particular, if
the source function $S_\nu$ is an even function of $(\nu-\nu_0)$, then
$u(\nu-\nu_0)$ will have an odd part only if $R(\nu,\nu')$ does. If
instead $R(\nu,\nu')=R(\nu',\nu)$, as holds for the redistribution
Cases I, II and III (Mihalas 1978, \S13-3) without atomic recoil
\footnote{Technically, redistribution Cases I, II and III do not allow
for atomic recoils. The modifications with recoils shall be designated
by adding the suffix 'r', thus, Case IIr.}, then $u(\nu-\nu_0)$ will
be even in $(\nu-\nu_0)$ and no energy exchange will occur. It is
evident from Eq.~(\ref{eq:Rdetbal}) that the asymmetry induced by
atomic recoil gives rise to the energy exchange in this case. These
symmetry relations are exploited below.

It is shown in the Appendix that under the assumption of Case IIr
redistribution (perfectly sharp lower state, finite radiative decay
time in upper state, frequency coherence in the restframe of the
atom), which is the case of most interest here, the integral over the
redistribution function may be expressed in terms of the Voigt line
profile $\varphi_V(\nu)$, so that
\begin{eqnarray}
G({\bf r},t)&=&\frac{\chi_0({\bf r},t)c}{h\nu_0}h
\int_0^\infty d\nu u_{\nu}({\bf r},t)\biggl[\epsilon\omega\varphi_V(\nu)
\nonumber\\
\phantom{\biggl[}&&-\frac{1}{2}\omega^2\frac{d\varphi_V(\nu)}{d\nu}\biggr],
\label{eq:GheatV}
\end{eqnarray}
where the effects of atomic recoil have been included to order
$\epsilon=h\nu_0/(2kTm_a c^2)^{1/2}$, where $T$ is the gas temperature
and $m_a$ is the mass of the recoiling atom, and $\omega=\nu_0b/c$,
where $b=(2kT/m_a)^{1/2}$ is the Doppler parameter. Because
$\varphi_V(\nu)$ is an even function of $(\nu-\nu_0)$,
Eq.~(\ref{eq:GheatV}) shows that energy transfer by recoil (the first
term in brackets) is driven by the even part of $u_\nu=u(\nu-\nu_0)$,
while energy transfer through the second term is driven by the odd
part. The odd part may also contain a term of order $\epsilon$, as it
will for an even source.

It is shown below that in the Fokker-Planck approximation, a source
function $S_\nu$ even in $(\nu-\nu_0)$ (or one well-approximated as
even, such as a function very smoothly varying across the resonance
frequency), will generate an odd part to $u(\nu-\nu_0)$ only through
recoils. Thus both terms in Eq.~(\ref{eq:GheatV}) will be proportional
to the recoil parameter $\epsilon$. It is also noted that any
non-negligible variation in frequency of the continuum source across
the resonant frequency may induce non-negligible energy transfer
between the radiation and matter even in the absence of
recoils. Expanding $S_\nu\approx S_{\nu_0}+(\nu-\nu_0)(dS/d\nu)_0$
shows that the correction term $(\nu-\nu_0)(dS/d\nu)_0$ will drive an
odd contribution to $u_\nu=u(\nu-\nu_0)$ in Eq.~(\ref{eq:unuev}) even
for a symmetric redistribution function ($R[\nu,\nu']=R[\nu',\nu]$),
and so create a non-vanishing energy exchange rate as described by
Eq.~(11).

Following Rybicki (2006), it is useful to recast
Eq.~(\ref{eq:GheatV}) in terms of the radiation colour temperature.
Motivated by Field's (1958) introduction of an effective colour
temperature of the radiation field, MMR presented a formal definition
of colour temperature $T_{\mathcal N}$ (called $T_\alpha$ in their paper) in
terms of the photon occupation number ${\mathcal N}(\nu)$:
\begin{equation}
T_{\mathcal N}=-\frac{h}{k}
\left[\frac{d\log{\mathcal N}(\nu)}{d\nu}\right]^{-1}.
\label{eq:TN}
\end{equation}
In order to account for stimulated scattering, Rybicki (2006) modified
the temperature definition to ${T^*_{\mathcal N}}(\nu)=T_{\mathcal
N}(\nu)[1+{\mathcal N}(\nu)]$.

In light of the form of Eq.~(\ref{eq:GheatV}), it is now suggested
that, in the absence of stimulated emission, the most natural
definition for the colour temperature for describing energy exchange
is in terms of the energy density of the radiation field. Through
straightforward manipulations, Eq.~(\ref{eq:GheatV}) may be recast in
the form
\begin{equation}
G=P_l n_l \frac{h\nu_0}{m_a c^2} h\nu_0 \left(1-\frac{T}
{\langle T_u\rangle_H}\right),
\label{eq:GheatVTu}
\end{equation}
where
\begin{equation}
P_l=\frac{B_{lu}}{4\pi}c \int_0^\infty\,d\nu \varphi_V(\nu) u_\nu
\label{eq:Pl}
\end{equation}
is the total photon scattering rate per atom in the lower state, and
the mean harmonic colour temperature has been defined by:
\begin{equation}
\langle T_u\rangle_H=\int_0^\infty\, d\nu u_\nu\varphi_V(\nu){\Bigg/}
\int_0^\infty\, d\nu u_\nu\varphi_V(\nu)\frac{1}{T_u(\nu)},
\label{eq:TuH}
\end{equation}
where
\begin{equation}
T_u(\nu)=-\frac{h}{k}\left(\frac{d\log u_\nu}{d\nu}\right)^{-1}.
\label{eq:Tu}
\end{equation}
The limit $T << \langle T_u\rangle_H$ of Eq.~(\ref{eq:GheatVTu})
recovers the heating rate given by MMR. At lower colour temperatures,
the heating rate is suppressed by the efficiency factor $1-T/\langle
T_u\rangle_H$.

The temperature in Eq.~(\ref{eq:TuH}) is identical to the ``light
temperature'' $T_L$ defined by Field (1958) in the context of the
Wouthuysen-Field mechanism, as will now be demonstrated. If the total
transition rates induced by the scattering of \Lya photons from the
lower hydrogen hyperfine $n=1$ level ($_0S_{1/2}$) in hydrogen to the
upper hyperfine level ($_1S_{1/2}$) is designated $P^L_{01}$, and the
reverse rate designated $P^L_{10}$, then Field (1958) (see also
Tozzi \etal 2000), shows that in equilibrium,
\begin{equation}
\frac{P^L_{01}}{P^L_{10}}=3\frac{\langle u_{\nu_0}\rangle +
\langle u_{\nu'_0}\rangle}{\langle u_{\nu_1}\rangle +
\langle u_{\nu'_1}\rangle},
\label{eq:hypereq}
\end{equation}
where $\langle u_{\nu_i}\rangle=\int\, d\nu\varphi_V(\nu-\nu_i)u_\nu$
is the energy density of the radiation field averaged over a Voigt
profile centred on $\nu=\nu_i$. The four frequencies $\nu_0$,
$\nu'_0$, $\nu_1$ and $\nu'_1$ correspond to the \Lya transitions
between the $n=1$ and $n=2$ pairs $(_0S_{1/2}, _1P_{1/2})$,
$(_0S_{1/2}, _1P_{3/2})$, $(_1S_{1/2}, _1P_{1/2})$, and $(_1S_{1/2},
_1P_{3/2})$, respectively. Field (1958) defines the effective colour
temperature $T_L$ of the radiation field by
$P^L_{01}/P^L_{10}=3(1-T_*/ T_L)$, where $kT_*=h\nu_{01}$ and
$h\nu_{01}$ is the energy difference between the $n=1$ hyperfine
levels. A straightforward expansion of the Voigt profiles about
$\varphi_V(\nu-\nu_0)$ to lowest order in $d\varphi_V/d\nu$ shows that
$T_L$ is identical to $\langle T_u\rangle_H$, further justifying this
as the most pragmatic definition of the colour temperature in the
absence of stimulated emission (cf Deguchi \& Watson 1985).

Rybicki (2006) expresses the heating rate in a form similar to
Eq.~(\ref{eq:GheatVTu}), except in terms of $T^*_{\mathcal N}(\nu)$
(his eq.~[32]), instead of $T_u(\nu)$. He then approximates the
scattering cross-section as a $\delta$-function to evaluate the colour
temperature at line-centre to obtain (his eq.~[35]):
\begin{equation}
G_{\rm Ryb}={P_l}^* n_l \frac{h\nu_0}{m_a c^2} h\nu_0 \left[1-\frac{T}
{{T^*_{\mathcal N}}(\nu_0)}\right].
\label{eq:GheatVTR}
\end{equation}
(In Rybicki's formulation, $u_\nu$ is approximated as $h\nu_0 n_\nu$
in the frequency-integral in the definition of $P_l$, and stimulated
emission is accounted for by multiplying the result by $[1+{\mathcal
N}(\nu)]$, the combined effect of which is defined here as ${P_l}^*$.)
The derivation here shows that the result is more general, and in
particular does not rely on the Fokker-Planck formulation. It does,
however, depend on the specific form Eq.~(\ref{eq:GheatV}).

The various definitions of colour temperature have relative
differences from $T_u$ of $kT_u/h\nu$. Provided $kT_u<<h\nu$, the
differences in these definitions are therefore small. Moreover,
because multiplicative factors of $\nu$ were dropped in adopting the
Voigt profile (so, for example, the Rayleigh scattering limit is not
recovered), the actual multiplicative factors of $\nu$ in defining the
radiation colour temperature are not accounted for in this
formalism. For these reasons, the approximation $u_\nu=h\nu_0 n_\nu$
is adopted below for the definition of the colour temperature for
convenience, to give $T_n(\nu)$ in analogy to Eq.~(\ref{eq:Tu}):\
\begin{equation}
\langle T_n\rangle_H=\int_0^\infty\, d\nu n_\nu\varphi_V(\nu){\Bigg/}
\int_0^\infty\, d\nu n_\nu\varphi_V(\nu)\frac{1}{T_n(\nu)},
\label{eq:TnH}
\end{equation}
where
\begin{equation}
T_n(\nu)=-\frac{h}{k}\left(\frac{d\log n_\nu}{d\nu}\right)^{-1}.
\label{eq:Tn}
\end{equation}

\subsection{Uniform expansion}

In a homogeneous and isotropic expanding medium, Eq.~(\ref{eq:unuev})
must be modified by adding expansion and redshifting terms to
$\partial u_\nu/ \partial t$. For this case, $Du_\nu/Dt$ becomes
(Peebles 1968; Peebles 1993),
\begin{equation}
\frac{Du_\nu}{dt}=\frac{\partial u_\nu}{\partial t}+3\frac{\dot R}{R}u_\nu
-\nu\frac{\dot R}{R}\frac{\partial u_\nu}{\partial \nu},
\label{eq:DunuDtexp}
\end{equation}
where $R(t)$ is the expansion factor. The equation may be simplified
further by defining $\tilde u_\nu =(R/R_0)^3u_\nu$ and $\tilde
S_\nu=(R/R_0)^3S_\nu$, where $R_0$ is the expansion factor at some
time $t_0$. In this case, the form of Eq.~(\ref{eq:unuev}) is
recovered if $u_\nu$ and $S_\nu$ are everywhere replaced by $\tilde
u_\nu$ and $\tilde S_\nu$, and $D{\tilde u}_\nu/Dt$ is now identified
with
\begin{equation}
\frac{D{\tilde u}_\nu}{Dt}=\frac{\partial {\tilde u}_\nu}{\partial t}
-\nu\frac{\dot R}{R}\frac{\partial {\tilde u}_\nu}{\partial \nu}.
\label{eq:DunuDtc}
\end{equation}
A similar result follows for the number density $n_\nu$ by first
expressing $u_\nu=h\nu n_\nu$ in Eq.~(\ref{eq:nnuev}), and then
replacing $n_\nu$ everwhere by $\tilde n_\nu=(R/R_0)^2n_\nu$,
$S_\nu$ by $\tilde S_\nu=(R/R_0)^2S_\nu$, and identifying $Dn_\nu/Dt$
with
\begin{equation}
\frac{D{\tilde n}_\nu}{Dt}=\frac{\partial {\tilde n}_\nu}{\partial t}
-\nu\frac{\dot R}{R}\frac{\partial {\tilde n}_\nu}{\partial \nu}.
\label{eq:DnnuDtc}
\end{equation}

An integration of the rate equation for $u_\nu$ exactly recovers the
heating rate of Eq.~(11), after expressing $Du_\nu/Dt$ by
Eq.~(\ref{eq:DunuDtexp}), which allows for the energy lost by the
radiation due to the expansion.

\subsection{Approximate formulations}
Several approximate form of the transfer equation for resonant
radiation exist in the literature. A common form starts from
Eq.~(\ref{eq:unuev}), but sets $h\nu=h\nu_0$ in the coefficient,
ignoring the frequency shift of a photon on scattering (Mihalas 1978,
\S2-1). The equation for the energy density $u_\nu$ then becomes
\begin{equation}
\frac{D u_\nu({\bf r},t)}{D t}
\cong \chi_0({\bf r},t) c
\left[\int_0^\infty d\nu' R(\nu',\nu)u_{\nu'}({\bf r},t)
-\varphi(\nu) u_\nu({\bf r},t)\right].
\label{eq:unuevap}
\end{equation}
While adequate for describing the evolution of $u_\nu$ for most cases,
this approximate form is inadequate for the purpose of computing the
rate of energy transfer between the photons and the scattering medium,
since the rate should be proportional to the frequency shift per
scattering event. A straightforward integration over frequency, noting
that $\int\,d\nu R(\nu',\nu)=\varphi(\nu')$, shows that this form for
the scattering equation gives a vanishing value for the net energy
exchange rate.

An intermediate approximation between the exact form and the
approximation above is provided by Basko (1978) for the photon number
density $n_\nu$. This starts from Eq.~(\ref{eq:nnuev}), but makes the
approximations $u_{\nu'}=h\nu_0n_{\nu'}$ and $u_\nu=h\nu_0n_\nu$ on
the right hand side to arrive at
\begin{equation}
\frac{D n_\nu({\bf r},t)}{D t}
\cong\chi_0({\bf r},t)c
\left[\int_0^\infty d\nu' R(\nu',\nu)n_{\nu'}({\bf r},t)
-\varphi(\nu) n_\nu({\bf r},t)\right].
\label{eq:nnuevap}
\end{equation}
Most Fokker-Planck treatments in the literature start with this form.
It has the advantage over Eq.~(\ref{eq:unuevap}) of recovering an
approximate form for the heating rate, given by multiplying by $h\nu$
and integrating over frequency
\begin{eqnarray}
G({\bf r},t)&=&\int_0^\infty d\nu S(\nu) - \frac{D}{Dt}
\int_0^\infty d\nu u_\nu({\bf r},t)\\
&\cong&\chi_0({\bf r},t)c
\int_0^\infty d\nu \int_0^\infty d\nu' (h\nu'-h\nu)
R(\nu',\nu)n_{\nu'}({\bf r},t).\nonumber
\label{eq:Gheatap}
\end{eqnarray}
It is identical to Eq.~(11) except that $u_{\nu'}$ has
been approximated by $h\nu_0n_{\nu'}$. Since the expression already
allows for a frequency shift of the scattered photon, however, this
approximation introduces only a small, and normally negligible, error.

\section{Fokker-Planck approximation}

\subsection{Basic equation}

The integro-differential character of the radiative transfer equation
with the full redistribution function makes its solution nearly
intractable, requiring special numerical methods. A useful technique
is to obtain solutions in a Fokker-Planck approximation. Basko (1978)
introduced a Fokker-Planck approximation for Eq.~(\ref{eq:nnuevap})
for an infinite homogeneous medium, although he included only the
dominant contribution to the line profile in the wings. Krolik (1990)
presents a formal expression for the Fokker-Planck approximation, but
similarly evaluates the coefficients only in the wings. Rybicki \&
Dell'Antonio generalised Basko's form by rederiving a Fokker-Planck
equation retaining the full Voigt line profile. Following Unno (1955)
and Harrington (1973), they expand $n_{\nu'}$ in
Eq.~(\ref{eq:nnuevap}) in a Taylor series about $\nu=\nu'$. Their
derived coefficients, however, did not explicitly conserve photon
number, and so they were forced to adjust the coefficients to ensure
the number of photons is conserved. An alternative derivation was
presented by Rybicki (2006) invoking detailed balance and stimulated
emission. It is still, however, based on a Taylor series expansion of
$n_\nu$.

A derivation of the Fokker-Planck equation is presented here which
explicitly conserves both particle number and energy while still
retaining the full line profile. For the present formulation, it is
useful to introduce the probability $W(\nu,Q)$ for scattering a photon
of frequency $\nu$ to one of frequency
$\nu'=\nu-Q$. Eq.~(\ref{eq:unuev}) may then be re-expressed as
\begin{eqnarray}
\frac{D u_\nu({\bf r},t)}{D t}
&=&\frac{\chi_0({\bf r},t)c}
{h\nu_0}h\nu\biggl[\int_{-\infty}^\infty dQ' W(\nu',Q')u_{\nu'}\nonumber\\
\phantom{\biggl[}&&-\int_{-\infty}^\infty dQ W(\nu,Q)u_\nu\biggr] +
h\nu S(\nu),
\label{eq:unuevw}
\end{eqnarray}
where $Q'=\nu'-\nu$, and $\varphi(\nu)=\int dQ W(\nu,Q)$ has been
used. It is also assumed here and elsewhere that $W(\nu,Q)$ becomes
vanishingly small for large $Q$, so that the lower integration bound
$\nu'=0$ in the first integral may be well-approximated by
$Q'\rightarrow-\infty$. Unlike previous approximations, the standard
Fokker-Planck approximation is based on a Taylor series expansion of
the product of both the scattering probability function and the
distribution function. A Taylor series expansion of
$W(\nu',Q')u_{\nu'}$ about $\nu'=\nu$ gives, to second order in
$\nu'-\nu$,
\begin{eqnarray}
W(\nu',Q')u_{\nu'}&\approx& W(\nu,Q')u_\nu + Q'\frac{\partial}
{\partial\nu}\left[W(\nu,Q')u_\nu\right]\nonumber\\
&& + \frac{1}{2}{Q'}^2
\frac{\partial^2}{\partial\nu^2}\left[W(\nu,Q')u_\nu\right].
\label{eq:Taylor}
\end{eqnarray}
Substituting this in Eq.~(\ref{eq:unuevw}) gives the canonical Fokker-Planck
form
\begin{eqnarray}
\frac{D u_\nu({\bf r},t)}{D t}
&\cong&\frac{\chi_0({\bf r},t)c}
{h\nu_0}h\nu\frac{\partial}{\partial \nu}\biggl\{\langle Q\rangle\varphi(\nu)
u_\nu({\bf r}, t)\\
&&+\phantom{\biggl\{}\frac{1}{2}\frac{\partial}{\partial \nu}
\left[\langle Q^2\rangle\varphi(\nu) u_\nu({\bf r}, t)\right]\biggr\}
+ h\nu S(\nu)\nonumber,
\label{eq:unuevFP}
\end{eqnarray}
where
\begin{equation}
\langle Q^n \rangle\varphi(\nu) \equiv
\left[\int_{-\infty}^\infty dQ  Q^n W(\nu,Q)\right],
\label{eq:Qpndef}
\end{equation}
and the dummy integration variable $Q'$ has been replaced by $Q$.
This form conserves energy to the accuracy of the approximation and to
order $b/c$, as may be shown by an integration by parts over $\nu$ and
noting that $W(\nu,Q)$ vanishes at the integration boundaries,
\begin{eqnarray}
G&=&\int_0^\infty d\nu h\nu S(\nu) - \frac{D}{Dt}\int_0^\infty d\nu
u_\nu({\bf r},t)\\
&=&\frac{\chi_0({\bf r},t)c}{h\nu_0}
\int_0^\infty d\nu h\langle Q \rangle\varphi(\nu) u_\nu({\bf r},t)\nonumber\\
&=&\frac{\chi_0({\bf r},t)c}{h\nu_0}\int_0^\infty d\nu \int_0^\infty d\nu'
(h\nu-h\nu') R(\nu,\nu')u_{\nu}({\bf r},t),\nonumber
\label{eq:GheatQ}
\end{eqnarray}
identical to Eq.~(11). This justifies estimating the
energy exchange rate between the photons and the gas using the
Fokker-Planck approximation, provided the approximation yields an
accurate solution for the radiation field.

Dividing Eq.~(30) through by $h\nu$ gives the
Fokker-Planck equation for the photon number density
\begin{eqnarray}
\frac{D n_\nu({\bf r},t)}{D t}
&=&\frac{\chi_0({\bf r},t)c}
{h\nu_0}\frac{\partial}{\partial \nu}\biggl\{\langle Q \rangle\varphi(\nu)
u_\nu\\
&&\phantom{\biggl\{}+\frac{1}{2}\frac{\partial}{\partial \nu}
\left[\langle Q^2 \rangle\varphi(\nu) u_\nu\right]\biggr\} + S(\nu),\nonumber
\label{eq:nnuevFP}
\end{eqnarray}
which is in explicit photon number conservation form (cf Krolik 1990).
The coefficients $\langle Q \rangle$ and $\langle Q^2 \rangle$ are
evaluated in the Appendix for Case IIr redistribution. For
$\epsilon=0$, they are identical to those derived by Rybicki \&
Dell'Antonio (1994) (noting that they expand in $-Q$ in the notation
here). They did not explicitly include the recoil terms, instead
adding by hand the result from Basko (1981). Here the recoil terms are
included self-consistently, resulting in additional higher order
corrections.

It is convenient to solve the equation in non-dimensional
form. Defining the dimensionless frequency shift
$x=(\nu-\nu_0)/\omega$ and the conformal time $\tau=\int_0^t dt
\chi_0({\bf r},t)c/\omega$, and substituting in the moments of $Q$ in
terms of the dimensionless Voigt profile
$\phi_V(x)=\varphi_V(\nu)\omega$, Eq.~(33) becomes
\begin{eqnarray}
\frac{Dn_x({\bf r},\tau)}{D\tau}&-&\frac{D\log\omega}{D\tau}n_x({\bf r},\tau)\\
&=&\frac{\partial}{\partial x}\left\{\left[\epsilon\phi_V(x)-\frac{1}{2}
\frac{d\phi_V(x)}{dx}\right]n_x({\bf r},t)\right\}\nonumber\\
&&+\frac{1}{2}\frac{\partial^2}{\partial x^2}\biggl\{\biggl[\phi_V(x)-
\frac{4}{3}\epsilon\frac{d\phi_V(x)}{dx}\nonumber\\
&&\phantom{\biggl\{}\phantom{\biggl[}+\frac{1}{3}\frac{d\phi_V^2(x)}{dx^2}
\biggr]n_x({\bf r},t)\biggr\}
+ \tilde S(x)\nonumber\\
&=&\frac{1}{2}\frac{\partial}{\partial x}\left[2\epsilon\phi_V(x)
n_x({\bf r},t)+\phi_V(x)\frac{\partial}{\partial x}n_x({\bf r},t)\right]
\nonumber\\
&&+\frac{1}{6}\frac{\partial^2}{\partial x^2}\left\{\left[-4\epsilon
\frac{d\phi_V(x)}{dx}+\frac{d^2\phi_V(x)}{dx^2}\right]n_x({\bf r},t)\right\}
\nonumber\\
&&+\tilde S(x)\nonumber\\
&=&\frac{1}{2}\frac{\partial}{\partial x}\biggl[2\epsilon\phi_V(x)
n_x({\bf r},t)+\phi_V(x)\frac{\partial}{\partial x}n_x({\bf r},t)\nonumber\\
&&\phantom{\biggl[}+\frac{2}{3}\epsilon\frac{d^2\phi_V(x)}{dx^2}n_x({\bf r},t)
+\frac{1}{3}\frac{d^2\phi_V(x)}{dx^2}\frac{\partial}{\partial x}n_x({\bf r},t)
\biggr]\nonumber\\
&&+\frac{1}{6}\frac{\partial}{\partial x}\biggl\{\left[-6\epsilon
\frac{d^2\phi_V(x)}{dx^2}+\frac{d^3\phi_V(x)}{dx^3}\right]n_x({\bf r},t)
\nonumber\\
&&\phantom{\biggl\{}-4\epsilon\frac{d\phi_V(x)}{dx}
\frac{\partial}{\partial x}n_x({\bf r},t)\biggr\}
+\tilde S(x),\nonumber
\label{eq:nxevFP}
\end{eqnarray}
where $n_x({\bf r},\tau)=n_\nu({\bf r},t)\omega$ and $\tilde
S(x)=S(\nu)\omega^2/(\chi_0 c)$ have been defined, a possible
time-evolution of $\omega$ has been allowed for, and terms of order
$xb/c$ have been neglected, consistent with the adoption of the Voigt
profile (see the Appendix). It should be noted that the valid recovery
of the effects of atomic recoil in this approximation requires $b/c <<
2\epsilon$. For hydrogen \Lya, this corresponds to
$T<<h\nu_0/k\approx10^5$K.

The approximations made by Rybicki \& Dell'Antonio (1994) correspond
to excluding the term in curly braces in the second equality of
Eq.~(34). The result has the form of a diffusion equation
with diffusion coefficient $\phi_V(x)/2$. The same form results from
the full Fokker-Planck equation when higher order derivatives of
$\phi_V(x)$ are neglected, as is appropriate far in the wings of the
profile (Harrington 1973; Basko 1978, 1981). It will be referred to
here as the Fokker-Planck diffusion approximation (FPDA), to
distinguish it from the standard Fokker-Planck approximation (FPA).

The formulation of Rybicki (2006) (his eq.~[20]), corresponds to
excluding the term in curly braces in the third equality of
Eq.~(34). It is another FPDA equation. It is adopted here
with the diffusion coefficient $D(x)=\phi_V(x)+(1/3)
d^2\phi_V(x)/dx^2$ (his eq.~[11]). The equation also includes an extra
term of order $b/c$ (arising from the term $-2n_\nu/\nu$ in his
eq.~[20]), and a quadratic term in $n_x$ arising from stimulated
emission. These latter two terms, not included in the discussion here,
play pivotal roles when the gas temperature approaches the equivalent
temperature of the resonant frequency or as the radiation thermalises
with the matter, which, in the case of an optically thick medium,
results in blackbody radiation. Comparisons with the approximation of
Rybicki (2006) are confined here to the diffusion equation terms using
the diffusion coefficient above.

\subsection{Higher order extension}
The above formalism is readily extended to higher orders. This is
useful when solving the equations to test the degree of convergence of
the original second-order Fokker-Planck approximation. Extended to
fourth order (by continuing the expansion in Eq.~(\ref{eq:Taylor}) to
fourth order), Eq.~(34) takes the form
\begin{eqnarray}
&&\frac{Dn_x({\bf r},\tau)}{D\tau}-\frac{D\log\omega}{D\tau}n_x({\bf r},\tau)\\
&=&\frac{\partial}{\partial x}\left\{\left[\epsilon\phi_V(x)-\frac{1}{2}
\frac{d\phi_V(x)}{dx}\right]n_x({\bf r},t)\right\}\nonumber\\
&+&\frac{1}{2}\frac{\partial^2}{\partial x^2}\left\{\left[\phi_V(x)-
\frac{4}{3}\epsilon\frac{d\phi_V(x)}{dx}+\frac{1}{3}\frac{d\phi_V^2(x)}{dx^2}
\right]n_x({\bf r},t)\right\}\nonumber\\
&+&\frac{1}{6}\frac{\partial^3}{\partial x^3}\biggl\{\biggl[\epsilon\phi_V(x)-
2\frac{d\phi_V(x)}{dx}+\frac{3}{2}\epsilon\frac{d\phi_V^2(x)}{dx^2}\nonumber\\
\phantom{\biggl\{}\phantom{\biggl[}&-&\frac{1}{4}\frac{d\phi_V^3(x)}{dx^3}
\biggr]n_x({\bf r},t)\biggr\}\nonumber\\
&+&\frac{1}{24}\frac{\partial^4}{\partial x^4}\biggl\{\biggl[4\phi_V(x)-
12\epsilon\frac{d\phi_V(x)}{dx}+3\frac{d\phi_V^2(x)}{dx^2}\nonumber\\
\phantom{\biggl\{}\phantom{\biggl[}
&-&\frac{8}{5}\epsilon\frac{d\phi_V^3(x)}{dx^3}+
\frac{1}{5}\frac{d\phi_V^4(x)}{dx^4}
\biggr]n_x({\bf r},t)\biggr\}\nonumber\\
&+& \tilde S(x).\nonumber
\label{eq:nxevFPho}
\end{eqnarray}
The third and fourth order terms correspond, respectively, to the
terms with third and fourth derivates of the expressions in curly
braces. Explicit expressions are provided in the Appendix for the
moments $\langle Q^3\rangle$ and $\langle Q^4\rangle$ from which the
higher order terms above are derived.

\subsection{Uniform expansion}

Eqs.~(34) and (35) continue to apply in an
isotropic and homogeneous expanding medium if $n_x$ is replaced by
${\tilde n}_x=(R/R_0)^2n_x$ and $D{\tilde n}_x/D\tau$ is given by
\begin{equation}
\frac{D{\tilde n}_x(\tau)}{D\tau}=
\frac{\partial {\tilde n}_x(\tau)}{\partial\tau}
-\gamma\frac{\partial{\tilde n}_x(\tau)}{\partial x},
\label{eq:DnxDtc}
\end{equation}
where
\begin{equation}
\gamma=\frac{\nu_0 H}{\chi_0 c}=\frac{8\pi}{3}\frac{H}{\lambda_0^3 n_l A_{ul}},
\label{eq:gamma}
\end{equation}
is the ratio of the scattering time at line-centre to the expansion
time $H^{-1}$ and is known as the Sobolev parameter (Chugai 1980;
Rybicki \& Dell'Antonio 1994). It is also the inverse of the optical
depth at line-centre $\lambda_0$ in a homogeneous and isotropic
expanding universe with Hubble parameter $H={\dot R}/R$ (Field 1959a).

\subsection{Method of solution}

It is instructive to express Eq.~(35) in Fourier
space. Denoting the Fourier transform of $n_x({\bf r},\tau)$ by
$\hat n_\kappa({\bf r},\tau)$, Eq.~(35) becomes
\begin{eqnarray}
\frac{D\hat n_\kappa({\bf r},\tau)}{D\tau}&-&\frac{D\log\omega}{D\tau}
\hat n_\kappa({\bf r},\tau)=\\
&&\frac{\kappa}{2\pi}\int_{-\infty}^\infty d\alpha\,\biggl\{
\biggl[\frac{1}{2}\alpha+\frac{1}{6}\kappa(\alpha^2-3)\nonumber\\
&&\phantom{\biggl[}+\frac{1}{24}\kappa^2\alpha(\alpha^2-8)+
\frac{1}{24}\kappa^3\left(4-3\alpha^2+\frac{1}{5}\alpha^4\right)\biggr]
\nonumber\\
&&-i\epsilon\biggl[1+\frac{2}{3}\kappa\alpha
+\frac{1}{4}\kappa^2\left(\alpha^2-\frac{8}{3}\right)\nonumber\\
&&\phantom{\biggl[}\phantom{\biggl[}+\frac{1}{2}\kappa^3\alpha
\left(\frac{2}{15}\alpha^2-1\right)\biggr]\biggr\}
\hat\phi_V(\alpha)\hat n_{\kappa-\alpha}({\bf r},\tau),\nonumber
\label{eq:nxevFPFTho}
\end{eqnarray}
where $\hat\phi_V(\alpha)$ is the Fourier Transform of $\phi_V(x)$.
Each order of $\kappa$ corresponds to the same order in the Taylor
series expansion. The difference between the second order form above
for the non-recoil terms and the Rybicki (2006) form is that in the
latter $\alpha/2$ above is replaced by $\alpha/2-\alpha^3/6$.

The equation is solved by fourth-order Runge-Kutta with an adaptive
timestep. The scheme is not fully stable for terms beyond the third
order. While adequate for producing the solutions presented in this
paper, it is not sufficiently efficient to integrate to much longer
times. Ultimately a modified Crank-Nicholson method would be most
desirable for solving Eq.~(35). Such an approach is
under investigation.

The heating rate expressed in Fourier space may be derived from
Eq.~(\ref{eq:GheatV}) and noting that
$\hat\phi_V(-\kappa)=\hat\phi_V(\kappa)$,
\begin{eqnarray}
G({\bf r},t)&=&\frac{\chi_0 c h}{2\pi}\int_{-\infty}^\infty d\kappa\,
\left(\epsilon-\frac{1}{2}i\kappa\right)\hat\phi_V(\kappa)\hat n_{\kappa}
({\bf r},\tau)\\
&=&\frac{\chi_0 c h}{2\pi}\int_{-\infty}^\infty d\kappa\,
\left[\epsilon{\hat n_\kappa}^{(r)}({\bf r},\tau)+
\frac{1}{2}\kappa{\hat n_\kappa}^{(i)}({\bf r},\tau)\right]
\hat\phi_V(\kappa),\nonumber
\label{eq:GheatFT}
\end{eqnarray}
where ${\hat n_\kappa}^{(r)}$ and ${\hat n_\kappa}^{(i)}$ are the real
and imaginary parts of $\hat n_\kappa$, noting that the real and
imaginary parts are, respectively, even and odd functions of $\kappa$
since $n_x$ is real. The heating rate may also be evaluated using
Eq.~(\ref{eq:GheatVTu}). To the accuracy of approximation in the
treatment here, $\langle T_u\rangle_H$ may be replaced by $\langle
T_n\rangle_H$. Expressed as a ratio of integrals in Fourier space,
\begin{equation}
\langle T_n\rangle_H=-\frac{h\omega}{k}\frac{\int_{-\infty}^\infty d\kappa\,
{\hat n_\kappa}^{(r)}{\hat \phi_V}(\kappa)}{\int_{-\infty}^\infty d\kappa\,
{\hat n_\kappa}^{(i)}\kappa{\hat \phi_V}(\kappa)}.
\label{eq:TnHF}
\end{equation}

Cosmological expansion is accounted for by identifying
$D{\hat n}_\kappa/D\tau$ with
\begin{equation}
\frac{D{\hat n}_\kappa(\tau)}{D\tau}=\frac{\partial {\hat n}_\kappa(\tau)}
{\partial\tau}+i\gamma\kappa {\hat n}_\kappa(\tau).
\label{eq:DnhatDtau}
\end{equation}
It is noted that even for an even source in $x$, cosmological
expansion will introduce an imaginary part to ${\hat n}_\kappa$. It
follows from Eq.~(39) that energy exchange will occur
even without atomic recoil. This results in a cooling term as the
photon distribution is redshifted through the line-profile (cf, Chen
\& Miralda-Escud\'e 2004). It is also noted that for a linear inflow
of matter, $\gamma$ changes sign and produces a heating term as
photons are blueshifted through the line-profile.

\section{Results}

\begin{figure}
\begin{center}
\leavevmode \epsfxsize=3.3in \epsfbox{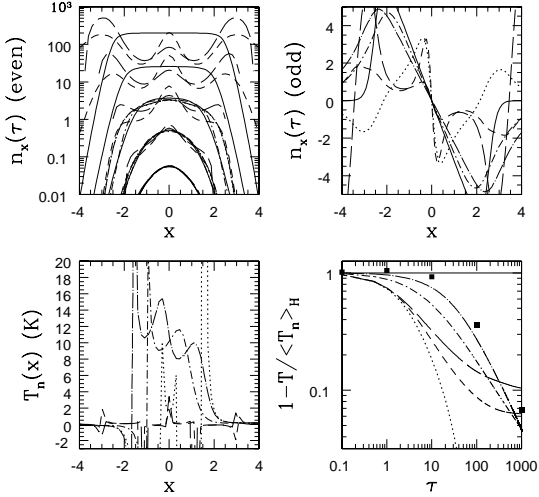}
\end{center}
\caption{Numerical solution of $n_x$ vs $x$ to the Field (1959b)
problem for a unit source with a Doppler profile. Shown are the
analytic solution (solid), the FPA solutions to second order
(short-dashed) and third (long-dashed) order, and the FPDA solutions
of Rybicki \& Dell'Antonio (1994) (dotted short-dashed) and Rybicki
(2006) (dotted long-dashed). Also shown is a truncated second order
FPA solution (dotted line; see text). The effect of atomic recoil is
included for a medium at $T=10$~K. {\it Top left panel}:\ The even
part of $n_x$ at the times, from bottom to top, $\tau=0.1$, 1, 10, 100
and 1000. Only the second and third order FPA solutions are shown at
$\tau=100$ and 1000 for clarity. In the absence of recoils, the even
part is identical to the full solution. {\it Top right panel}:\ The
odd part of $n_x$. The odd part vanishes in the absence of atomic
recoil. {\it Bottom left panel}:\ The specific temperature $T_n(x)$ at
$\tau=1000$. {\it Bottom right panel}:\ The evolution of the mean
harmonic temperature $\langle T_n\rangle_H$ in terms of the heating
efficiency. The points correspond to the approximation
Eq.~(\ref{eq:GheatVTR}) applied to the Rybicki (2006) solution.
}
\label{fig:FieldS}
\end{figure}

Field (1959b) derived an analytic solution to the problem of the
scattering of resonant photons in a homogeneous and isotropic medium
with a Doppler profile source term. The problem is not particularly
well-suited to the Fokker-Planck approximation because of the rapidly
varying source function, but it does provide a test of the
approximation and its convergence. Solutions to
Eq.~(35) are found with $S(x)=\exp(-x^2)/\pi^{1/2}$ for
four cases:\ the second and third order FPA equations, and the FPDA
equations of Rybicki \& Dell'Antonio (1994) and Rybicki (2006).  The
results for cold hydrogen gas at $T=10$K are shown in
Figure~\ref{fig:FieldS}. The solutions are separated into their even
and odd parts, as, for a source even in $x$, the effect of atomic
recoil appears only in the odd part of the solution, which is
proportional to $\epsilon$. (The contribution of recoils to the even
part is of order $\epsilon^2$.) The original Rybicki \& Dell'Antonio
approximation best matches the analytic solution near the
line-centre. The solution of Rybicki (2006) agrees well, but not quite
as well as the original Rybicki \& Dell'Antonio (1994) form.

By contrast, the second order Fokker-Planck approximation derived here
shows appreciable features. The subtraction of $\alpha^3/6$ from
$\alpha/2$ in Eq.~(38) in the Rybicki (2006)
formulation plays the critical role of eliminating the oscillatory
behaviour. This corresponds to the highest (fourth order) derivative
of $\phi_V(x)$ in Eq.~(34). Truncating the equation at
the third derivative of $\phi_V(x)$ by subtracting this term from the
second order solution eliminates the oscillatory
behaviour. Accordingly, a truncated second order FPA (TFPA) equation
is solved with this term removed. The term, however, is removed only
from the equation for the {\it real} part of ${\hat n}_\kappa$, not
the imaginary, for the reason explained below. Although the (full)
third order Fokker-Planck solution agrees more closely with the
analytic solution than the second order, the oscillatory behaviour is
still present.

Field (1959b) demonstrated that when atomic recoil is included, the
radiation field reaches statistical equilibrium with the matter near
the line-centre, in the sense that for $\epsilon \vert x\vert<<1$, the
radiation field asymptotically approaches the form $n_x \sim
n_x^{(e)}[1-2\epsilon x]\simeq n_x^{(e)}\exp[-h(\nu-\nu)/kT]$ for
$\tau>>1$. It follows from Eq.~(\ref{eq:Tn}) that the colour
temperature at line-centre is $T_n(0)=T$. Since the analytic
correction to the profile due to recoil is proportional to $x$, the
extent over which the profile assumes this form is unclear. Field
(1959b) argued it should apply within a flattened core with the
time-asymptotic value $n_x(\tau)\simeq \tau/[2(\log\tau)^{1/2}]$ at
line-centre, at frequencies smaller than a critical frequency
$x_c=(\log\tau)^{1/2}$. In Figure~\ref{fig:FieldS}, the odd part of
the analytic solution is plotted as
$n_x^{(o)}=n_x^{(e)}[\exp(-2\epsilon x)-\exp(2\epsilon x)]/2$, noting
that it is strictly applicable only for $\epsilon \vert
x\vert<<1$. The slow logarithmic growth of the critical frequency
suggests the radiation field may reach statistical equilibrium outside
the Doppler core only very slowly. Moreover, the value of the colour
temperature depends also on how flat the core is away from
line-centre. This suggests the heating rate may not become vanishingly
small for many scattering times at line-centre ($\tau>>1$).

Figure~\ref{fig:FieldS} shows that the frequency-specific colour
temperature $T_n(\nu)$ approaches the gas temperature at line-centre
only for the Rybicki \& Dell'Antonio (1994) and Rybicki (2006)
approximations. Even for these cases, $T_n(\nu)$ deviates considerably
from the gas temperature well within the core of the line profile. In
the other cases, the colour temperature takes on mostly very small
values, both positive and negative. The lesson would appear to be that
this temperature has little physical meaning, except possibly very
near the line-centre. The colour temperature in the Fokker-Planck
approximation, however, appears to converge to the equilibrium
temperature at line-centre only very slowly with increasing
order. Much away from line-centre, it is not even clear the colour
temperature should converge to the gas temperature, except for the
extreme blackbody limit.

As discussed in the previous section, both the heating rate and the
Wouthuysen-Field efficiency depend on the mean harmonic temperature
$\langle T_n\rangle_H$. Weighting by the line profile produces a mean
harmonic temperature that evolves toward the matter temperature for
large $\tau$, as shown in Figure~\ref{fig:FieldS}.

It is noted that while subtracting $\alpha^3/6$ from $\alpha/2$ in the
equation for the real part of $\hat n_\kappa$ in
Eq.~(38) removes the oscillations in the second order
FPA solution, subtracting this term from the equation for the
imaginary part as well produces a solution that results in a rapid,
and unphysical, decline in $\langle T_n\rangle_H$ to values well below
the gas temperature. Subtracting the term only from the real part
yields $\langle T_n\rangle_H\rightarrow9.8\, {\rm K}\simeq T$ by
$\tau\gsim100$, and stays constant at this value up to
$\tau=1000$. The light temperature has reasonably converged to the
matter temperature within the error of the approximation. The
truncated version of the second order FPA equation thus appears a
viable approximation for including the effects of atomic recoil while
recovering the correct solution to the radiative transfer equation, in
spite of the lack of a mathematical justification.

The form of $T_n(\nu)$ in the different approximations has a curious
property. While it agrees nearly exactly with the gas temperature in
the Rybicki \& Dell'Antonio (1994) and Rybicki (2006) Fokker-Planck
diffusion approximations at line centre, this is no longer true in the
full Fokker-Planck approximations. In these cases, the cancellations
among the negative and positive values of $T_n(\nu)$ ensure the mean
harmonic average colour temperature $\langle T_n\rangle_H$ converges
to the gas temperature. For the Rybicki \& Dell'Antonio (1994),
Rybicki (2006) and truncated second order Fokker-Planck
approximations, the convergence is rapid, with the efficiency factor
scaling roughly like $1-T/\langle T_n\rangle_H\sim \tau^{-1/2}$ for
$\tau>>1$. For the other cases, the convergence starts rapidly, but by
$\tau>100$ levels off at efficiency values of $5-10$\% by
$\tau=1000$. It is unclear how large $\tau$ must be before the
efficiency declines to 1\% and heating essentially shuts off. Going to
third order indeed suggests the time is longer than given by the
second order Fokker-Planck approximation. The integration method used
is too inefficient to integrate to much longer times or to higher
orders, so that it is not possible to answer this question here.

The points in the lower right hand panel of Figure~\ref{fig:FieldS}
correspond to Eq.~(\ref{eq:GheatVTR}) applied to the Rybicki (2006)
solution. At early times this results in efficiency factors exceeding
unity, which is unphysical. At late times, the approximation agrees
reasonably well with the Rybicki (2006) predicted efficiency.

\begin{figure}
\begin{center}
\leavevmode \epsfxsize=3.3in \epsfbox{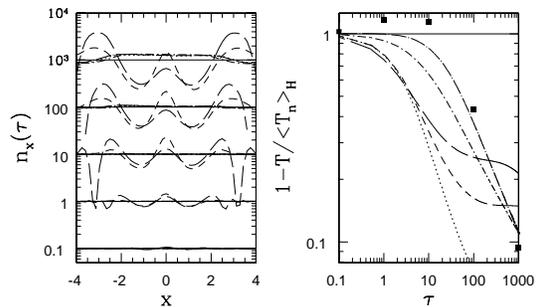}
\end{center}
\caption{Numerical solution of $n_x$ vs $x$ for a constant continuum
source. A Voigt scattering profile is adopted with a gas temperature
$T=10$~K, corresponding to $a=0.015$.  Shown are the no-scattering
solution (solid), the FPA solutions to second order (short-dashed) and
third order (long-dashed), and the FPDA solutions of Rybicki \&
Dell'Antonio (1994) (dotted short-dashed) and Rybicki (2006) (dotted
long-dashed). Also shown is a truncated second order FPA solution
(dotted line; see text). The effect of atomic recoil is included for a
medium at $T=10$~K. {\it Left panel}:\ The evolution of $n_x$ at the
times, from bottom to top, $\tau=0.1$, 1, 10, 100 and 1000.  {\it
Right panel}:\ The evolution of the mean harmonic temperature $\langle
T_n\rangle_H$ in terms of the heating efficiency. The points
correspond to the approximation Eq.~(\ref{eq:GheatVTR}) applied to the
Rybicki (2006) solution.
}
\label{fig:CS_RII}
\end{figure}

The case of a constant continuum source $S(x)=1$ is shown in
Figure~\ref{fig:CS_RII}. The gas temperature is $T=10$~K,
corresponding to $a=0.015$ for the Voigt profile. The second and third
order FPA solutions again show substantial structure, and the
efficiencies reach plateaus for $\tau>100$ at values of $15-20$\% by
$\tau=1000$. Again, heating persists at a non-negligible level for
many scattering times at line centre. For the two FPDA and the TFPA
solutions, the heating efficiency continues to decline, again scaling
roughly as $\tau^{-1/2}$ for $\tau>>1$. The points in the right hand
panel correspond to the approximation Eq.~(\ref{eq:GheatVTR}) for the
efficiency applied to the Rybicki (2006) solution. The values are
again unphysically high at early times.

\section{Discussion \& Conclusions}

\subsection{Fokker-Planck approximations}

Previous formulations of the time-dependent Fokker-Planck
approximation to the radiative transfer equation for the scattering of
resonant photons were based on a Taylor series expansion of only the
radiation intensity in the scattering integral (Unno 1955; Harrington
1973; Rybicki \& Dell'Antonio 1994; Rybicki 2006). The result of the
approximation has the form of a diffusion equation. In this paper, an
equation based on a Taylor series expansion of the product of the
scattering probability and the intensity is obtained, which is the
standard Fokker-Planck approximation. This permits an easier analysis
of the improved accuracy of the approximation to higher orders. In
this paper, the resulting approximation has been derived for Case II
scattering using the full Voigt profile, and including the first-order
contributions from atomic recoil. The approximation correctly
conserves particle numbers with well-defined coefficients, unlike the
Fokker-Planck diffusion approximation.

Time-dependent solutions are presented to the FPDA and FPA equations
for the case of a unit Doppler profile source (for Case I) and for a
constant continuum source. An additional term in the second order FPA
equation introduces artificial features within the cores of the
profiles. These diminish in the third order, but are still
pronounced. Truncating the second order FPA equation by removing this
term eliminates the features. In principle, the features should be
eliminated by going to a sufficiently high order in the Fokker-Planck
expansion. Such an investigation, however, requires an alternative
integration method to that used here.

\subsection{Approach to statistical equilibrium}

An expression for the heating rate of the scattering medium through
atomic recoil is derived. It is shown that the Fokker-Planck
approximation correctly recovers the full heating rate to the accuracy
of the approximation. Both the heating rate and the effectiveness of
the Wouthuysen-Field mechanism depend on the mean harmonic colour
temperature $\langle T_u\rangle_H$ computed from the
frequency-specific temperture $T_u(\nu)$, weighted by the absorption
profile. For photons of frequency $\nu$ in statistical equilibrium
with the scattering medium at temperature $T$, it has usually been
expected that $T_u(\nu)$ converges to $T$ for large $\tau$, where
$\tau$ is the time expressed in units of the scattering time at
line-centre. This was shown to be the case analytically by Field
(1959b) for two particular problems involving Case I redistribution,
at least near the line centre. Numerical solutions are found to
Field's problem for a pure Doppler profile source. Using the FPDA
approximations of Rybicki \& Dell'Antonio (1994) and Rybicki (2006),
$T_u(\nu)\rightarrow T$ is found at the line centre for $\tau>>1$, but
away from the line centre $T_u(\nu)$ deviates appreciably from $T$,
and even takes on negative values. For the FPA equations, $\vert
T_u(\nu)\vert$ is generally small compared with $T$, and takes on both
negative and positive values. The third order FPA equation predicts
the larger value at line centre. Presumably higher order Fokker-Planck
approximations must yield the correct temperature at the
line-centre. The investigation of this question is in progress.

The more physically-motivated quantity $\langle T_u\rangle_H$ is found
to converge toward $T$. The rate of convergence, however, is sensitive
to the form of the approximation. The most rapid convergence is found
for the truncated Fokker-Planck approximation.  Whether or not this is
an artifact of the form of the approximation is unclear. Convergence
is also found for the Fokker-Planck diffusion approximations. For the
full Fokker-Planck approximation developed here, the convergence of
$\langle T_u\rangle_H$ to $T$ substantially slows for $\tau>100$. Going
to third order results in an even slower convergence rate.

\subsection{Angular dependence}

The results presented have all assumed an isotropic radiation field.
In situations in which the gas is heated by one or more local sources,
the incident radiation field may be highly anisotropic. One approach
is to take angular moments of the radiative transfer
equation. Neglecting stimulated emission, the upper level population
in equilibrium is given by
\begin{equation}
n_u(\nu)A_{ul} = n_l\frac{B_{lu}}{4\pi}\oint
\frac{d\omega'}{4\pi} \int d\nu'\,
R(\nu',{\bf {\hat n}'};\nu,{\bf \hat n})I_{\nu'}({\bf r},t,{\bf {\hat n}'}),
\label{eq:nudbang}
\end{equation}
and the emissivity becomes
\begin{equation}
\eta_\nu=h\nu n_l\frac{B_{lu}}{4\pi}\oint
\frac{d\omega'}{4\pi} \int d\nu'\,
R(\nu',{\bf {\hat n}'};\nu,{\bf \hat n})I_{\nu'}({\bf r},t,{\bf {\hat n}'}).
\label{eq:etanuang}
\end{equation}
The angular coupling between the redistribution function and the
intensity precludes a simple form for the energy transfer rate. A
solution to the radiative transfer equation along these lines is
sketched by Rybicki \& Dell'Antonio (1994).

\subsection{Astrophysical implications}

The Wouthuysen-Field mechanism for coupling the spin temperature to
the colour temperature of resonant photons has been discussed in a
variety of contexts, including the IGM, intergalactic gas clouds like
Damped Lyman-Alpha Absorbers, gas on the peripheries of galaxies, and
in clouds near quasars, with applications not only to ${\rm ^1H}$, but
${\rm ^2D}$ and ${\rm ^3He^+}$ as well (eg, Field 1958; Deguchi \&
Watson 1985). The mechanism may be particularly complicated for ${\rm
^3He^+}$ because of the \Lya radiation from ${\rm ^4He^+}$ and the
coincidence of the ${\rm ^3He^+}$ \Lya transition with a Bowen
fluorescence line of \OIII. The reader is referred to Deguchi \&
Watson (1985) for a discussion of these and other complicating
effects. The discussion in this section is confined to more general
remarks.

The scattering times at line-centre for ${\rm ^1H}$, ${\rm ^2D}$ and
${\rm ^3He^+}$, normalised to the hydrogen number density $n_{\rm H}$
(in ${\rm cm}^{-3}$) using the Big Bang Nucleosynthesis constrained
abundances of ${\rm ^2D}$ and ${\rm ^3He}$ from Burles, Nollett \&
Turner (2001), are
\begin{eqnarray}
t_{\rm sca}&=&
3.2\, n_{\rm H}^{-1}T^{1/2}~{\rm s}\quad \qquad; {\rm ^1H},\nonumber\\
&=&7.3\times10^4\, n_{\rm H}^{-1}T^{1/2}~{\rm s}\quad ; {\rm ^2H},\nonumber\\
&=&1.0\times10^4\, n_{\rm H}^{-1}T^{1/2}~{\rm s}\quad ; {\rm ^3He^+}.
\label{eq:tsca}
\end{eqnarray}
These timescales are typically short for neutral gas. For ionized gas,
however, as for instance in Lyman Limit Systems, the rates may become
quite long, particularly for ${\rm ^2D}$ and ${\rm ^3He^+}$. For
instance, in a cloud with $n_{\rm HI}\approx10^{-5}\,{\rm cm}^{-3}$ at
$T=10^4$K, the timescale is about $10^{12}\,{\rm s}$ for these
isotopes. If it takes 10--100 scattering times for the
Wouthuysen-Field mechanism to bring the colour temperature $\langle
T_u\rangle_H$ to the gas temperature, the equilibrium timescale
approaches that of the lifetime of massive stars. If such objects, or
their \HII regions, are the sources of the \Lya photons, it may be
that while the colour temperature at the \HI resonance has come into
equilibrium, that of \DI has not, in which case an erroneous ${\rm
D/H}$ ratio may be inferred from the comparison of radio measurements
of the two hyperfine transitions. Conversely, if the cosmic ${\rm
D/H}$ ratio is accepted, such observations may provide an estimate of
the time since the radiation sources turned on. Even for hydrogen, the
colour temperature may not reach equilibrium if the source varies on a
timescale that does not greatly exceed the scattering time.

\begin{figure}
\begin{center}
\leavevmode \epsfxsize=3.3in \epsfbox{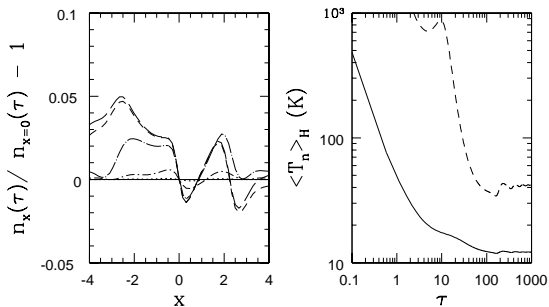}
\end{center}
\caption{Numerical solution of $n_x$ vs $x$ for a constant continuum
source in an expanding medium with $\gamma=0.1$ A Voigt scattering
profile is adopted with a gas temperature $T=10$~K, corresponding to
$a=0.015$. The results for the truncated FPA solution are shown in the
left panel at the times $\tau=0.1$ (dotted), 1 (dot short-dashed), 10
(dot long-dashed), 100 (short-dashed) and 1000 (long-dashed). The
photons are increasingly redshifted with time. The light temperature
$\langle T_n\rangle_H$ is shown in the right panel for the truncated
second order FPA equation (solid line) and the Rybicki (2006) FPDA
equation (dashed line).
}
\label{fig:CCS_cold}
\end{figure}

\begin{figure}
\begin{center}
\leavevmode \epsfxsize=3.3in \epsfbox{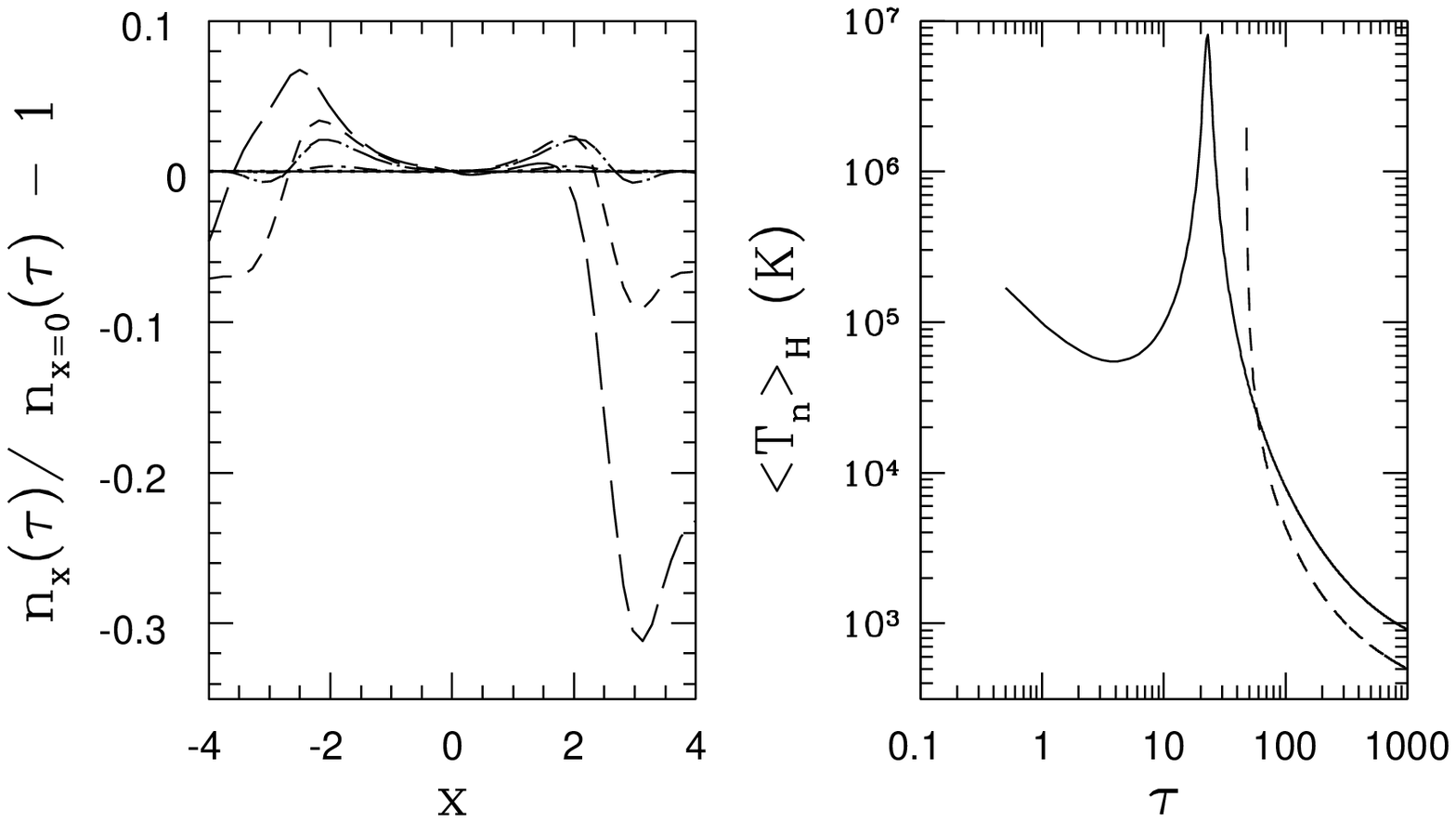}
\end{center}
\caption{Numerical solution of $n_x$ vs $x$ for a constant continuum
source in an expanding medium with $\gamma=0.01$.  A Voigt scattering
profile is adopted with a gas temperature $T=2\times10^4$~K,
corresponding to $a=3.3\times10^{-4}$. The results for the truncated
FPA solution are shown in the left panel at the times $\tau=0.1$
(dotted), 1 (dot short-dashed), 10 (dot long-dashed), 100
(short-dashed) and 1000 (long-dashed). The photons are increasingly
redshifted with time. The light temperature $\langle T_n\rangle_H$ is
shown in the right panel for the truncated second order FPA equation
(solid line) and the Rybicki (2006) FPDA equation (dashed line).
}
\label{fig:CCS_hot}
\end{figure}

Cosmological expansion will generally play a negligible role in the
approach to statistical equilibrium of the radiation field. This is
because the smallness of the Sobolev parameter. Rybicki \& Dell'Antonio
(1994) show that, within the damping wings, expansion negligibly affects
the photon distribution until a time $\tau\approx (a/\gamma^4)^{1/3}>>1$
for typical values of $a$ and $\gamma$ expected for the IGM.

Expansion, however, may play a role in more local situations, such as
the outflow from a star or an active galactic nucleus (AGN). One
possible effect is a freeze-out of $\langle T_u\rangle_H$. For the
sake of demonstrating the principle, a computation with $a=0.015$ and
$\gamma=0.1$ is show in Figure~\ref{fig:CCS_cold} for cold hydrogen
gas at $T=10$K. The photon distribution is skewed to the red by the
expansion. Both the truncated second order FPA solution and the
Rybicki (2006) FPDA solution predict a freeze-out of $\langle
T_u\rangle_H$ above the gas temperature for $\tau>100$, but at
different values. The radiation density profiles differ slightly near
line centre, with the dip at $x=0$ in Figure~\ref{fig:CCS_cold} not as
deep in the FPDA solution.

For hot ionised hydrogen gas, the expansion may result in a light
temperature well below the matter temperature. This is illustrated in
Figure~\ref{fig:CCS_hot} for $a=3.3\times10^{-4}$ and $\gamma=0.01$
for gas at $T=2\times10^4$K. The photon distribution is strongly
skewed toward the red by $\tau=1000$, with a sharp cut-off in the blue
similar to the Chugai (1980) steady-state limiting solution for the
radiative transfer in the profile wings of photons from a
monochromatic source at line-centre in an expanding medium. Both the
truncated second order FPA solution and the Rybicki (2006) FPDA
solution predict $\langle T_u\rangle_H<T$ for $\tau\gsim60$. The
Rybicki (2006) FPDA predicts negative values of $\langle T_u\rangle_H$
for $\tau\lsim50$. It should be noted, however, that at the assumed
temperature the requirement $2\epsilon<<b/c$ is only marginally
met. The full Rybicki (2006) formulation includes an extra term of
order $b/c$ which is not incorporated into the formalism here. This
term may eliminate the negative values, although it is noted that
negative values are still found in a similar integration using
$T=1000$K.

The possible heating role played by atomic recoil was originally
proposed by MMR as a possible means of heating the IGM at high
redshifts before it is re-ionised. Given the rapidity with which
$\langle T_u\rangle_H\rightarrow T$, this now appears unlikely (cf,
Chen \& Miralda-Escud\'e 2004; Rybicki 2006), particularly if the
predictions using the Rybicki \& Dell'Antonio (1994), Rybicki (2006)
or truncated second order Fokker-Planck approximations are most
accurate. The prediction of the FPA equations, however, is a levelling
off of the efficiency to values of 10-20\% for large $\tau$. This
suggests that heating may persist for $\tau>>1$. How long is still
unclear. The delay may also be an artifact of the artificial
oscillatory behaviour of the solutions.

The heating rate of MMR at full efficiency may be expressed as
\begin{equation}
G_{\rm MMR}=\frac{h\nu_0}{m_a c^2}h\nu_0 n_l P_l.
\label{eq:GMMR}
\end{equation}
The amount of energy per atom in the lower level that may deposited
before the efficiency becomes vanishingly small may be expressed as
\begin{equation}
\Delta E=f (G_{\rm MMR}/ n_l)t_{\rm sca}=f\frac{h\nu_0}{m_a c^2}
\frac{\omega {\bar n}_{\nu_0}}{n_l}h\nu_0,
\label{eq:DeltaE}
\end{equation}
where ${\bar n}_{\nu_0}=\int\, d\nu\phi_V(\nu)n_\nu$. For the hydrogen
\Lya transition, $h\nu_0/m_ac^2\approx10^{-8}$ and $h\nu_0\gsim10^5$K.
Consequently a very large ratio of photons within a Doppler width of the
resonance to neutral hydrogen atoms is typically required for an appreciable
increase in temperature. Expressed in terms of the average photon occupation
number ${\bar{\mathcal N}}(\nu_0)$ at line-centre,
\begin{eqnarray}
\Delta E&=&f\frac{h\nu_0}{m_a c^2}\frac{b}{c}
\frac{8\pi{\bar{\mathcal N}}(\nu_0)}{n_l\lambda_0^3}h\nu_0\nonumber\\
&\approx&8\times10^{-14}\left(\frac{f}{10}\right)T^{1/2}n_l^{-1}
\left[\frac{{\bar{\mathcal N}}(\nu_0)}{10^{-21}}\right]\, {\rm K},
\label{eq:DeltaEN0}
\end{eqnarray}
where the expression has been evaluated for hydrogen \Lya in the last
line, and normalised to ${\bar{\mathcal N}}(\nu_0)=10^{-21}$, corresponding
to a mean intensity of ${\bar J}_{\nu_0}=2.2\times10^{-22}\,{\rm erg\, cm^{-2}
\, s^{-1}\, Hz^{-1}\, sr^{-1}}$. Thus in a normal intergalactic environment,
the effect is completely negligible even prior to the reionization epoch.

There may be some circumstances, however, where the effect may contribute
non-negligibly to the total heating. For instance, for blackbody radiation
at $T=10^4$K, ${\bar{\mathcal N}}(\nu_0)\approx\exp[-h\nu_0/kT]\approx8\times
10^{-6}$. For a somewhat dilute blackbody not in thermal equilibrium with
the matter, recoils may provide a significant amount of heating. Whether or
not it would be competitive with other processes depends on the particular
situation.

At the opposite extreme, cold tenuous gas near a luminous source such
as a young star or active galactic nucleus, if shielded from
ionizing radiation, could be heated by several degrees or more.

Another situation in which recoil heating may play a role is in an
\HII region. According to the on-the-spot approximation, the photon
occupation number at line centre due to diffuse recombination radiation
is (Osterbrock 1974),
\begin{equation}
{\mathcal N}_d(\nu_0)=\left(\frac{h^2}{2\pi m_ekT}\right)^{3/2}
\frac{n_{\rm HII}n_e}{n_{\rm HI}}\approx4\times10^{-22}T_4^{-3/2}
\frac{1}{\chi^2_{\rm HI}}n_{\rm HI},
\label{eq:Nd}
\end{equation}
where $T_4$ is $T$ in units of $10^4$K, $\chi_{\rm HI}$ is the neutral
hydrogen fraction, and $m_e$ is the mass of an electron. Using this in
Eq.~(\ref{eq:DeltaEN0}) gives
\begin{equation}
\Delta E\approx3\times10^{-12}\left(\frac{f}{10}\right)T_4^{-1}
\frac{1}{\chi^2_{\rm HI}}\, {\rm K}.
\label{eq:DeltaEchi2}
\end{equation}
It is useful to express this in terms of a source of ionising photons $S$
for a \Lya optical depth $\tau_{\rm HI}$ through a region of size $R$. Then
\begin{equation}
\Delta E\approx 3 S_{50}{\tau_{\rm HI}}^{-1}{R_{\rm pc}}^{-1}
T_4^{-1}\, {\rm K},
\label{eq:DeltaES}
\end{equation}
where $S_{50}$ is in units of $10^{50}\, {\rm s^{-1}}$ and
$R_{\rm pc}$ is in units of parsecs. The scattering of diffuse \Lya
photons may provide a significant source of heat in the development of
an \HII region near an ultraluminous star or an AGN.

\bigskip
\section*{acknowledgments}

The author thanks Peter Brand for helpful conversations.


\begin{appendix}

\section{Moments of the redistribution function}

The moments $\langle Q\rangle\varphi(\nu)$ of the redistribution
probability $W(\nu,Q)$ are computed here. These results generalize
those of Rybicki \& Dell'Antonio (1994) to explicitly take into
account atomic recoil. Accordingly, the results are derived here from
first principles.

Consider the absorption of a photon of frequency $\nu$ travelling in
direction ${\bf \hat n}$ and re-emission of a photon of frequency $\nu'$
travelling in direction ${\bf \hat n'}$ by an atom travelling with velocity
${\bf v}$ in the laboratory frame. It is useful to project ${\bf v}$ onto
an orthogonal basis given by the average and difference of the photons
directions (Hummer 1962):
\begin{equation}
{\bf \hat n_1}=\gamma_+({\bf\hat n + \hat n'}),\qquad
{\bf \hat n_2}=\gamma_-({\bf\hat n - \hat n'}),\qquad
{\bf \hat n_3}={\bf\hat n \times \hat n'},
\label{eq:basis}
\end{equation}
where $\gamma_\pm=1/[2(1\pm\mu)]^{1/2}$ and $\mu={\bf \hat n\cdot\hat
n'}$. The corresponding components of the atom velocity are $(v_1,
v_2, v_3)$. To lowest order in $v/c$ and $h\nu_0/m_a c^2$, where $m_a$
is the mass of the recoiling atom, the frequency shift on scattering
is given by
\begin{equation}
\nu'-\nu=\nu_0\left[\frac{\bf v}{c}\cdot({\bf \hat n' -\hat n})-
\frac{h\nu_0}{m_a c^2}(1-\mu)\right].
\label{eq:recoil}
\end{equation}

Allowing for radiation damping in the upper level, the absorption profile
is given by the Lorentz profile in the restframe of the atom
\begin{equation}
f(\nu)=\frac{\delta/\pi}{\left[(\nu-\nu_0)^2+\delta^2\right]},
\label{eq:Lorentz}
\end{equation}
where $\nu_0$ is the line-centre frequency and $\delta=\Gamma/4\pi$,
where $\Gamma$ is the radiative damping width of the upper state. In
expression Eq.~(\ref{eq:Lorentz}), the implicit approximation
$\nu/\nu_0\approx1$ is made, except for the frequency shift in the
denominator, and so does not recover the correct behaviour far from
line centre, such as the Rayleigh limit (Peebles 1969; Peebles
1993). Since this paper is concerned solely with energy transfer
resulting from resonant scattering, this is an adequate
approximation. Radiation heat exchange at other frequencies must be
dealt with separately.

It is assumed that the atomic velocites are distributed as a Maxwellian
with temperature $T$ according to
\begin{equation}
P(u)d^3u=\pi^{-3/2}\exp(-u^2)d^3u,
\label{eq:Maxwell}
\end{equation}
where ${\bf u}={\bf v}/b$, and $b=(2kT/m_a)^{1/2}$ is the Doppler
parameter. It is useful to introduce the Doppler width $\omega=\nu_0
b/c$, frequency offset $x=(\nu-\nu_0)/\omega$, recoil parameter
$\epsilon=(h\nu_0/m_a c^2)c/b=h\nu_0/(2kTm_a c^2)^{1/2}$, and
normalised damping constant $a=\delta/\omega$. The frequency shift
may then be expressed as
$x-x'=\gamma_{-}^{-1}u_2+\epsilon(1-\mu)$. Then the probability that
an incident photon of frequency shift $x$ is scattered to $x'=x-q$ is
given by
\begin{eqnarray}
w(x,q,\mu)&=&\frac{a}{\pi^2}\int_{-\infty}^\infty du_1
\int_{-\infty}^\infty du_2\exp(-u_1^2-u_2^2)\\
&&\times\frac{\delta[q-\gamma_-^{-1}u_2-\epsilon(1-\mu)]}
{a^2 + \left[x-\gamma_+u_1(1+\mu)-\gamma_-u_2(1-\mu)\right]^2},\nonumber
\label{eq:wxqpm}
\end{eqnarray}
neglecting terms of order $xb/c$ in the denominator.

It is noted that the form of $w(x,q,\mu)$ satisfies the asymmetry
relation Eq.~(\ref{eq:Rdetbal}) in the limit $h\nu_0>>kT$ and
$\nu'/\nu\approx1$ (to order $xb/c$). The reverse scattering
of $x\rightarrow x'=x-q$ is $x'=x-q\rightarrow x'- (-q)$. The
corresponding relation between $w(x,q,\mu)$ and $w(x-q,-q,\mu)$.
Expressing $w(x-q,-q,\mu)$ according to Eq.~(A5) and
making the change of variable $u_2={\tilde u}_2-2\gamma_- q$ shows
after straightforward manipulations that
\begin{eqnarray}
w(x-q,-q,\mu)&=&e^{-2\epsilon q}w(x,q,\mu)\nonumber\\
&=&\exp\left[-\frac{h(\nu-\nu')}{kT}\right]w(x,q,\mu).
\label{eq:wdetbal}
\end{eqnarray}

The angle-averaged coefficients $\langle Q^n\rangle\varphi(\nu)$ may be
expressed in terms of $w(x,q,\mu)$ according to
\begin{equation}
\langle Q^n\rangle\varphi(\nu)=\frac{1}{2}\omega^{n-1}\int_{-1}^{1}
d\mu a_n(x,\mu),
\label{eq:Qpnanxm}
\end{equation}
where
\begin{equation}
a_n(x,\mu)=\int_{-\infty}^{\infty}dq q^n w(x,q,\mu).
\label{eq:anxmw}
\end{equation}
Following Rybicki \& Dell'Antonio (1994), such integrals are most
easily evaluated in Fourier space. From Eq.~(A5), the
Fourier transform of $w(x,q,\mu)$ is
\begin{eqnarray}
\hat w(\kappa,\lambda,\mu)&=&\int dx\int dq \exp(ix\kappa)\exp(iq\lambda)\nonumber\\
&&\frac{a}{\pi^2}\int_{-\infty}^\infty du_1
\int_{-\infty}^\infty du_2\exp(-u_1^2-u_2^2)\nonumber\\
&&\times\frac{\delta[q-\gamma_-^{-1}u_2-\epsilon(1-\mu)]}
{a^2 + \left[x-\gamma_+u_1(1+\mu)-\gamma_-u_2(1-\mu)\right]^2}\nonumber\\
&=&\exp\biggl[-a|\kappa|-\frac{1}{4}\kappa^2+i\lambda\epsilon(1-\mu)\nonumber\\
\phantom{\biggl[}&&-\frac{1}{2}\lambda\kappa(1-\mu)-
\frac{1}{2}\lambda^2(1-\mu)\biggr].
\label{eq:FTw}
\end{eqnarray}
The Fourier transform of $a_n(x,\mu)$ may be expressed as
\begin{equation}
\hat a_n(\kappa,\mu)=\frac{\partial^n}{\partial(i\lambda)^n}
\hat w(\kappa,\lambda,\mu)\Bigg\vert_{\lambda=0},
\label{eq:FTaFTw}
\end{equation}
noting that factors of $q^n$ correspond to $n$ derivatives with respect to
$i\lambda$ in Fourier space. The first five values of $\hat a_n(\kappa,\mu)$
are then
\begin{eqnarray}
\hat a_0(\kappa,\mu) &=&\exp\left[-a\vert\kappa\vert - \frac{1}{4}\kappa^2\right];\\
\hat a_1(\kappa,\mu) &=&\left[\epsilon(1-\mu)+\frac{1}{2}i\kappa(1-\mu)\right]
\nonumber\\
&&\times\exp\left[-a\vert\kappa\vert - \frac{1}{4}\kappa^2\right];\\
\hat a_2(\kappa,\mu) &=&\left[(1-\mu)+(1-\mu)^2\left(i\epsilon\kappa-
\frac{1}{4}\kappa^2\right)\right]
\nonumber\\
&&\times\exp\left[-a\vert\kappa\vert - \frac{1}{4}\kappa^2\right];\\
\hat a_3(\kappa,\mu) &=&\biggl[3(1-\mu)^2\left(\epsilon+\frac{1}{2}i\kappa
\right)\nonumber\\
&&\phantom{\biggl[}-\frac{1}{8}(1-\mu)^3\left(6\epsilon\kappa^2
+i\kappa^3\right)\biggr]\nonumber\\
&&\times\exp\left[-a\vert\kappa\vert - \frac{1}{4}\kappa^2\right];\\
\hat a_4(\kappa,\mu) &=&\biggl[3(1-\mu)^2+6(1-\mu)^3\left(i\epsilon\kappa-
\frac{1}{4}\kappa^2\right)\nonumber\\
&&\phantom{\biggl[}+\frac{1}{16}(1-\mu)^4\left(-8i\epsilon\kappa^3
+\kappa^4\right)\biggr]\nonumber\\
&&\times\exp\left[-a\vert\kappa\vert - \frac{1}{4}\kappa^2\right].
\label{eq:aFT}
\end{eqnarray}

The $n=0$ term corresponds to the full absorption profile
$\varphi(\nu)$. Since for Case II redistribution, this corresponds to
the Voigt profile, the notation $\varphi_V(\nu)$ is adopted for
$\varphi(\nu)$. Its dimensionless form will be denoted as
$\phi_V(x)$. The Fourier transform of $\phi_V(x)$ is given by $\hat
a_0(\kappa,\mu)$ in Equation~(A11). The first five moments of $Q$
are then
\begin{eqnarray}
\langle Q^0\rangle &=& 1;\\
\langle Q\rangle &=& \epsilon\omega-\frac{1}{2}\omega^2
\frac{d\log\varphi_V(\nu)}{d\nu};\\
\langle Q^2\rangle &=& \omega^2-\frac{4}{3}\epsilon\omega^3
\frac{d\log\varphi_V(\nu)}{d\nu}\nonumber\\
&&+\frac{1}{3}\omega^4
\frac{1}{\varphi_V(\nu)}\frac{d^2\varphi_V(\nu)}{d\nu^2};\\
\langle Q^3\rangle &=& 4\epsilon\omega^3-2\omega^4
\frac{d\log\varphi_V(\nu)}{d\nu}+\frac{3}{2}\epsilon\omega^5
\frac{1}{\varphi_V(\nu)}\frac{d^2\varphi_V(\nu)}{d\nu^2}\nonumber\\
&&-\frac{1}{4}\omega^6
\frac{1}{\varphi_V(\nu)}\frac{d^3\varphi_V(\nu)}{d\nu^3};\\
\langle Q^4\rangle &=& 4\omega^4-12\epsilon\omega^5
\frac{d\log\varphi_V(\nu)}{d\nu}+3\omega^6
\frac{1}{\varphi_V(\nu)}\frac{d^2\varphi_V(\nu)}{d\nu^2}\nonumber\\
&&-\frac{8}{5}\epsilon\omega^7
\frac{1}{\varphi_V(\nu)}\frac{d^3\varphi_V(\nu)}{d\nu^3}\nonumber\\
&&+\frac{1}{5}\omega^8
\frac{1}{\varphi_V(\nu)}\frac{d^4\varphi_V(\nu)}{d\nu^4}.
\label{eq:Qmom}
\end{eqnarray}

\end{appendix}

\end{document}